\documentclass[
    ,final            
  ]
  {aipproc}

\layoutstyle{6x9}

\usepackage{setspace}
\usepackage{times}
\usepackage{graphicx}

\def\gtorder{\mathrel{\raise.3ex\hbox{$>$}\mkern-14mu
 \lower0.6ex\hbox{$\sim$}}}
\def\ltorder{\mathrel{\raise.3ex\hbox{$<$}\mkern-14mu
 \lower0.6ex\hbox{$\sim$}}}

\def\ge{G_E}
\def\gm{G_M}
\def\gep{G_{Ep}}
\def\gmp{G_{Mp}}

\def\gmn{G_{Mn}}
\def\mugegm{\mu_p G_E / G_M}
\def\gegm{G_E / G_M}

\begin{document}

\title{New measurements of the proton's size and structure using polarized
photons}



\author{J. Arrington}{address={Physics Division, Argonne National Laboratory, Argonne, IL 60439}}

\begin{abstract}

Improved measurements of the proton's structure are now possible thanks to
significant technical advances that allow us to probe the proton with
\textit{polarized} photons. These measurements have shown that the proton is
not as simple as previously believed: quark orbital angular momentum and
relativistic effects play an important role and the spatial distribution of
charge and magnetization do not simply mimic the spatial distribution of the
quarks. Even more recently, the large scale structure and size of the proton 
have been examined more carefully, and a significant discrepancy has been
observed between the charge radius of the proton as measured in the Lamb shift
of muonic hydrogen and measurements using the electron--proton interaction.

\end{abstract}

\maketitle


While the proton and neutron are thought of as the basic building blocks of
visible matter, they are, in fact, complicated bound states of quarks and
gluons, held together by the strong interactions described by Quantum
Chromodynamics (QCD). The forces binding quarks together are so powerful that
it is impossible to simply pluck a single quark from a proton; attempting to
do so requires energies so large that new quarks and gluons are ``ripped'' out
of the vacuum, forming new bound states (hadrons) around the quark one is
attempting to isolate.  Therefore, studying the interactions of QCD involves
careful examination of the internal structure of \textit{bound states} of
quarks to isolate information that is directly sensitive to the underlying
quark degrees of freedom.  Because of this, the proton plays an important dual
role as both a basic building block of visible matter and the most accessible
bound state of QCD, and as such, has been a primary focus for generations of
nuclear physicists.

In the last several years, new experimental tools have led to unprecedented
improvements in our ability to study the internal structure of the proton and
neutron.  Measurements utilizing polarized electron beams allow the proton to
be probed with polarized virtual photons, and the comparison of spin-flip
and non-spin-flip interactions provides access to small and difficult to
measure components of the protons electromagnetic structure. These
measurements have been made possible by advances in high-current,
high-polarization electron beams along with improved polarized targets and
recoil polarimeters.  These techniques are now in common usage at Jefferson Lab
(JLab) after pioneering work at SLAC, BATES, and Mainz~\cite{perdrisat07,
arrington07a, arrington11a}.

The distributions of the proton's charge and magnetization are encoded in the
elastic electromagnetic form factors, which are measured in elastic
electron--proton scattering.  The elastic e--p scattering cross section
depends on the two elastic form factors, $\ge(Q^2)$ and $\gm(Q^2)$, which
depend only on the square of the four-momentum transfer, and which describe
the difference between scattering from a structureless particle and an
extended, complex object.  By measuring the form factors, we probe the spatial
distribution of the proton charge and magnetization, providing the most direct
connection to the spatial distribution of quarks inside the proton.  In the
non-relativistic limit, the form factors are simply Fourier transforms of the
charge and magnetization spatial densities.  The size of the proton is related
to the strength of the forces that bind quarks and gluons together, while the
detailed distributions provide information on the quark dynamics, and yield
strong constraints on models of the proton internal structure.

\begin{figure}[t]
\includegraphics[angle=0,height=3.3in]{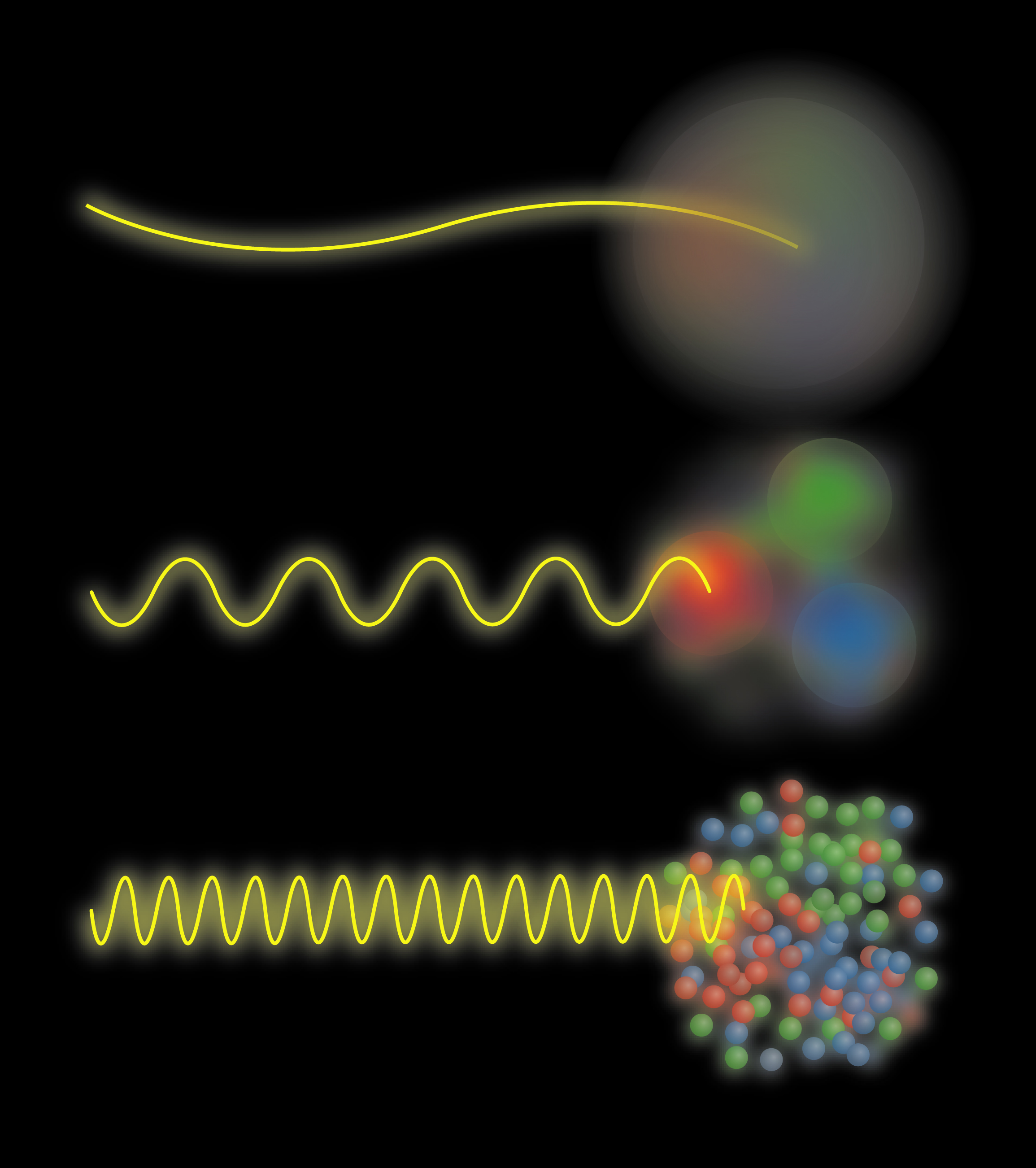}\\
\includegraphics[angle=0,height=3.3in]{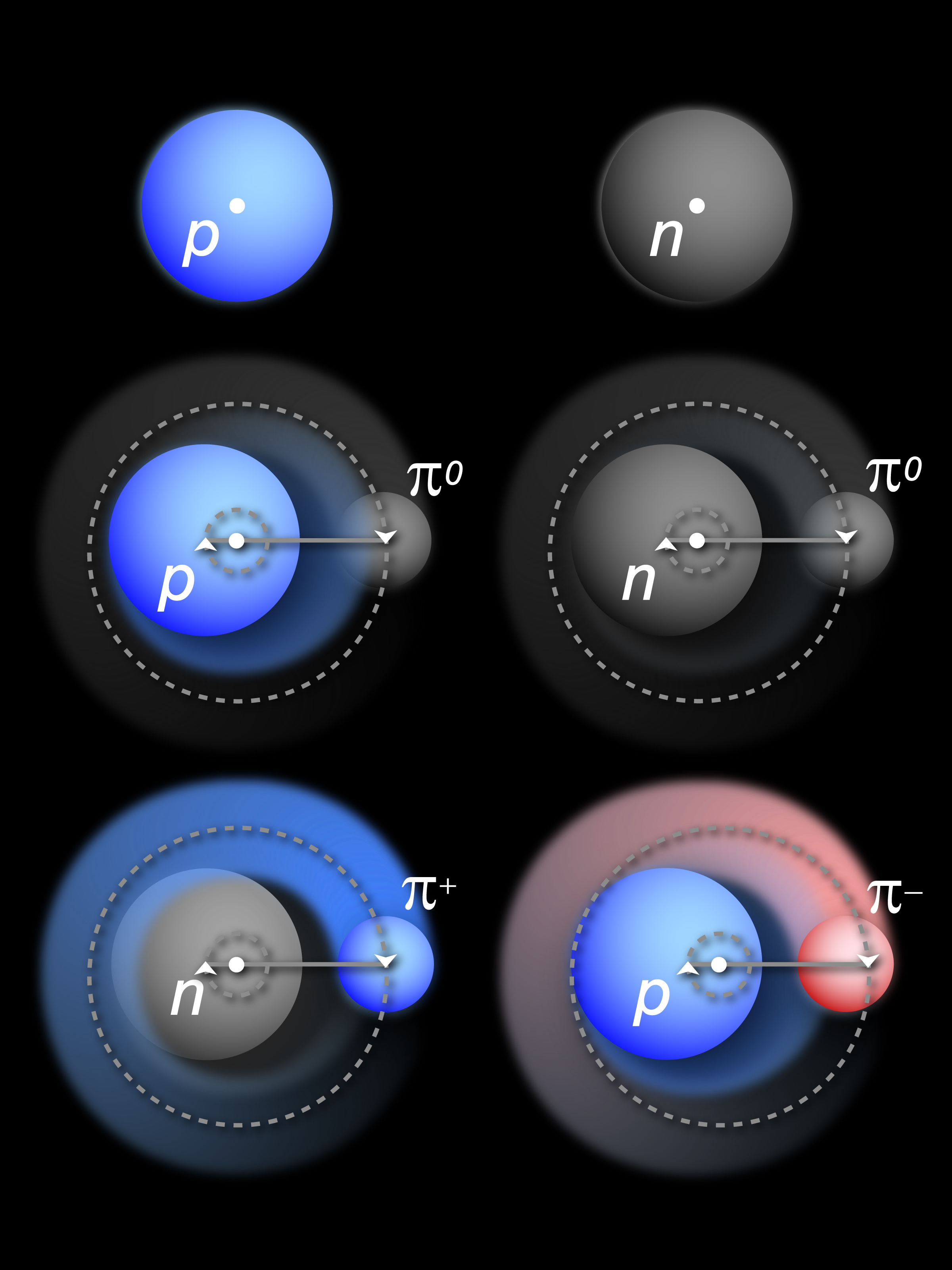}
\caption{\textbf{Left:} Illustration of the structure probed in
electron--proton scattering at various energy scales. At low energy, the
virtual photon probes the proton at large distance scales and is sensitive to
the size and large-scale structure.  At higher energies, the photons probe
shorter distance scales, probing the constituent quarks or the sea of
quark-antiquark pairs. \textbf{Right:} Illustration of the impact of the
``pion cloud'' to the charge distribution of the proton (left) and neutron
(right) charge radii. Blue (red) indicates positive (negative) charge, and
presence of a virtual pion--nucleon has two main effects.  First, the motion
of the system yields a 'smearing' of the nucleon charge distribution due to
its center-of-mass motion.  Second, the charge pion associated with the n
$\to$ p + $\pi^-$ fluctuation yields a negative contribution at large
distances. \textit{Credit: Joshua Rubin, Argonne National Laboratory}}
\label{fig:scale}
\end{figure}

In electron scattering, the electron interacts with the proton via exchange
of a virtual photon, characterized by its momentum transfer, $Q^2$.  At low
$Q^2$, the long wavelength photon is sensitive to the size and large scale
structure of the proton, while at high $Q^2$, the photon probes the fine
details of the structure, as illustrated in Figure~\ref{fig:scale} (left
panel). In the Born (single photon exchange) approximation, the elastic e--p
cross section is proportional to the reduced cross section, $\sigma_R =
[(Q^2/4m^2) \gm^2(Q^2) + \varepsilon \ge^2(Q^2)]$, where $m$ is the proton
mass, $\theta$ is the electron scattering angle, and
$\varepsilon^{-1}=1+2(1+Q^2/4m^2)\tan^2(\theta/2)$. The presence of the
angle-dependent factor $\varepsilon$ multiplying $\ge$ allows a Rosenbluth
separation~\cite{rosenbluth50} of the form factors by examining the
$\varepsilon$ dependence of $\sigma_R$ at fixed $Q^2$.

For over 50 years, extractions of the form factors using the Rosenbluth
technique appeared to show that the normalized charge and magnetic form factors
were approximately equal for the proton, and were similar to the magnetic form
factor of the neutron. This is consistent with non-relativistic quark models
of the proton and neutron structure, where the nucleon's charge and
magnetization are simply carried by the quarks, with identical spatial
distributions for up and down quarks. The only deviation from
this simple picture was the small, non-zero result for the electric form factor
of the neutron.  This indicated a small difference between the up and
down quark distributions, since identical spatial densities for
up and down quarks would yield perfect cancellation of their
charges at all points in space, resulting in a vanishing electric form factor.

Because the cross section is proportional to a combination of two terms, it is
difficult to extract one form factor when the other dominates the cross
section, limiting extractions of $\gm$ at very low $Q^2$ and $\ge$ at
very high $Q^2$. This difficulty led to interest in polarization
measurements which depend only on the \textit{ratio} $\gegm$, but which
require polarized electron beams and either polarized targets or proton recoil
polarimeters. Combining cross section and polarization measurements allows a
clean extraction of both form factors, even where one is strongly suppressed
in the cross section. With the new polarization techniques, it has become
possible to study in much greater detail the charge and magnetization
distributions.  This allows us to look for deviations from our simple picture,
and to more carefully examine contributions that yield differences between
up- and down-quark distributions, such as the ``pion cloud''
of the nucleon~\cite{Fried}, illustrated in Fig.~\ref{fig:scale} (right
panel).  With the inclusion of parity violating elastic scattering, it is also
possible to isolate the contribution of strange quarks to the proton
form factors~\cite{beck01,armstrong12}, given sufficiently precise knowledge
of the up- and down-quark contributions to make a reliable
extraction of the strange-quark contribution.

\begin{figure}[htbp]
\includegraphics[angle=0,width=.48\textwidth]{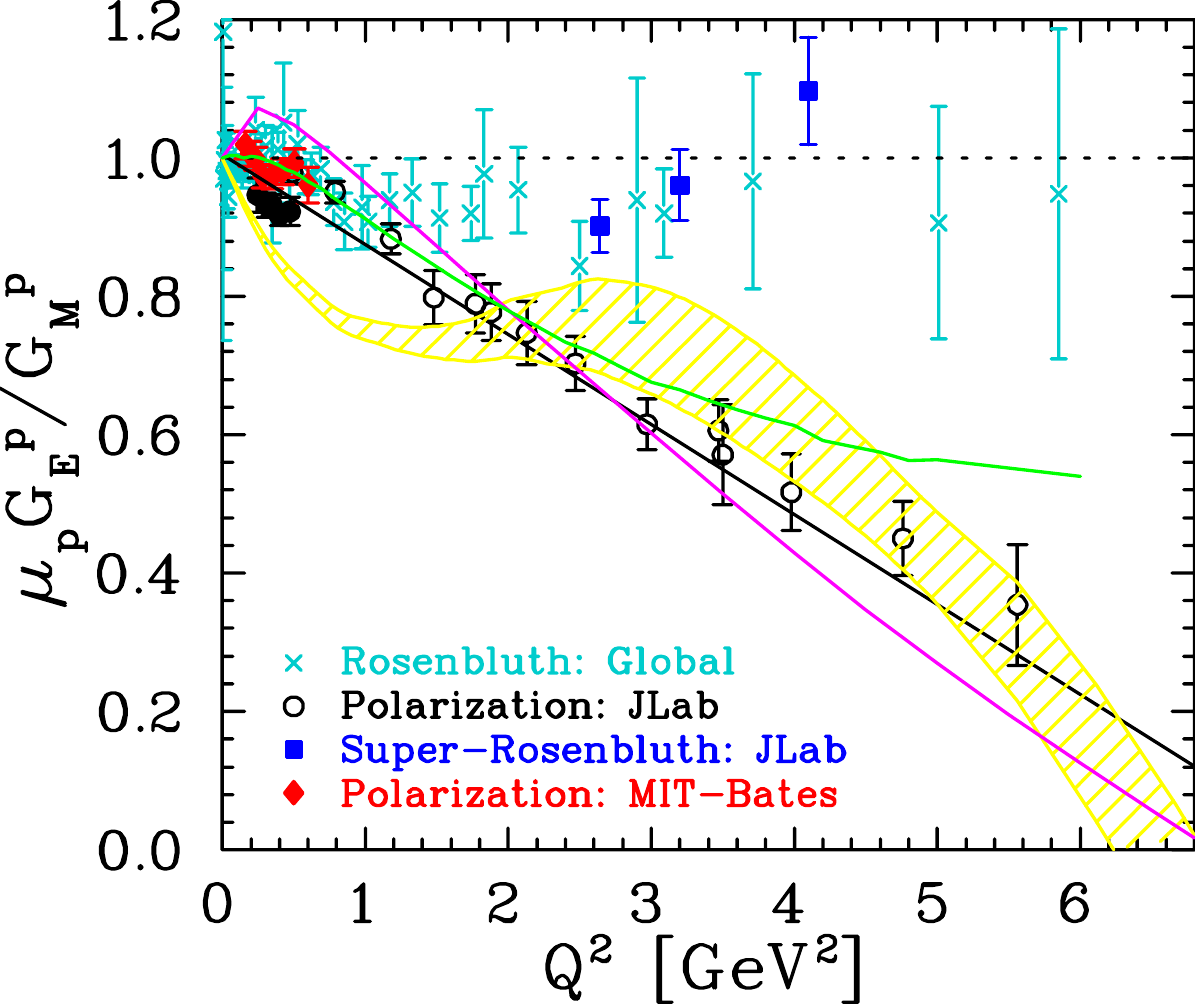}
\includegraphics[angle=0,width=.48\textwidth]{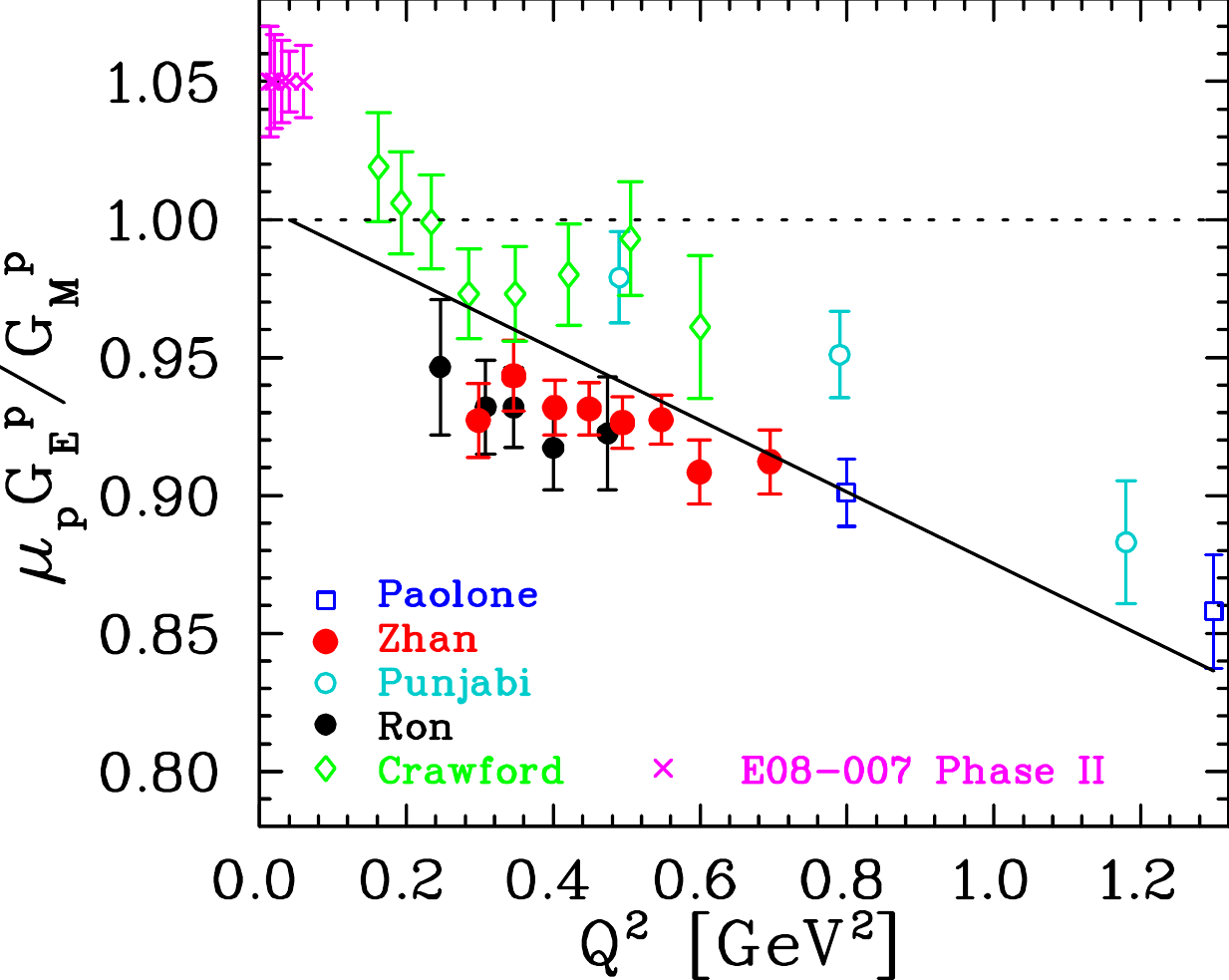}
\caption{\textbf{Left:} The proton form factor ratio $\mugegm$ from Rosenbluth
measurements~\cite{arrington03, christy04, qattan05} (neglecting TPE
corrections) and polarization measurements~\cite{punj, gayou, blast,
puckett10, puckett12, zhan11, ron11, paolone10}. The black curve is the fit
from Ref~\cite{arrington07c}, the magenta curve is the calculation of
Ref.~\cite{frank96}, the green curve is the calculation of 
Ref.~\cite{cardarelli00}, and the yellow band is the prediction of
Ref.~\cite{holl05}. The precise polarization data show a rapid drop with
$Q^2$, while the Rosenbluth data are, on average, consistent with unity.
{Right:} Polarization measurements of the ratio at low $Q^2$ along with the
projected uncertainties for Phase II of JLab E08-007, which recently 
completed data taking.}
\label{fig:gegm}
\end{figure}

One of the biggest surprises of the JLab form factor program was the falloff
of $\gep/\gmp$ at high $Q^2$~\cite{punj,gayou,puckett10}, which was at odds with
decades of cross section measurements~\cite{arrington03}. The high-$Q^2$
polarization data indicates that the ratio of electric to magnetic form
factors is constant at very low $Q^2$, then falls almost linearly with $Q^2$,
as shown in Figure~\ref{fig:gegm}.  This contradicted previous measurements
which showed $\mugegm \approx 1$, in agreement with the simple non-relativistic
model.  The difference between the polarization and cross section extractions is now
believed to be the result of two-photon exchange contributions.  These have
little impact on polarization measurements but significantly affect Rosenbluth
extractions of $\ge$ at high $Q^2$, making it appear as though $\mugegm
\approx 1$~\cite{guichon03, arrington04a, arrington07c, arrington11b}. Definitive
tests of the effect on the cross sections using precise comparisons of positron
and electron scattering are currently under way~\cite{arrington09}.

While this decrease of $\mugegm$ with $Q^2$ was largely unexpected, it is
perhaps not so surprising that relativistic effects become important at these
large momenta, and that the predictions of non-relativistic models break down.
These results led to an explosion of theoretical interest in trying to
understand the proton form factors~\cite{perdrisat07, arrington07a}.  The
difference in the $Q^2$ dependence of $\gep$ and $\gmp$ suggests a significant
difference in the charge and magnetization densities of the
proton~\cite{kelly02}, although one avoids model-dependent corrections only
when working in the infinite-momentum frame~\cite{miller07, miller08,
miller09a}.  While calculations which reproduce the falloff of $\gegm$ explain
the effect in somewhat different terms~\cite{frank96, cardarelli00, miller02a,
boffi, holl05},
most of these link the effect directly or indirectly to significant
contributions from quark orbital angular momentum~\cite{belitsky03, brodsky04a,
ralston04}. In addition, interpretation of these results in terms of
generalized parton distributions led to studies of the correlation between the
spatial distribution of the quarks and the spin or momentum that they carry,
showing that the spherically symmetric distribution of the proton is formed
from a rich collection of complex overlapping structures~\cite{miller07,
miller08, miller09a, miller03, belitsky04, carlson08b}.

Recent measurements of the neutron charge form factor at $Q^2$ values up to
3.4~GeV$^2$~\cite{riordan10} allow for more detailed examination of the up- and
down-quark contributions~\cite{CJRW2011}, showing significant
difference between the quark contributions to $F_1$ and $F_2$.  For the up
quarks, the form factors appear to fall as $Q^{-4}$ above $Q^2$=1.5~GeV$^2$,
while the down quarks decrease as $Q^{-2}$.  One interpretation of this is
the importance of the quark-diquark structure of the proton, with the more
strongly bound scalar (u--d) diquarks yielding a significant difference in the
high-$Q^2$ behavior of the up- and down-quark
contributions~\cite{cloet09, cloet2012}.  The combination of precise
measurements on the proton and the neutron at high-$Q^2$, including recent
measurements of $\gmn$~\cite{lachniet09}, also allows us to test models of the
nucleon against a complete set of form factor measurements covering a wide
range of momentum transfers.  As different models make varied assumptions
about the most relevant symmetries and degrees of freedom of QCD, this
improves our ability to identify the aspects of the underlying physics that
are most critical in determining the proton structure.  Such studies will be
further extended with the JLab 12 GeV upgrade~\cite{dudek12}, with the
additional advantage that higher $Q^2$ studies will be less sensitive to
pion cloud contributions, which are often added to the three-quark core
in more simplified approximations.

As illustrated by the examples above, the measurements at large momentum
transfer have provided significant new information on the
details of the proton structure, including the importance of relativity, quark
angular momentum, and the flavor structure of the proton.  By going to lower
momentum transfer and thereby utilizing longer wavelength probes, we become
sensitive to the size of the proton.  This is the region where one expects to
see the impact of the ``pion cloud'' of the proton, which arises from brief
fluctuations of the proton into a bound system of a proton or neutron and
pion, illustrated in Fig.~\ref{fig:scale}.  Because of the nucleon-pion mass
difference, the nucleon will be located nearer the center of mass, and the
pion will contribute at large distances. So the proton--$\pi^0$ contribution
yields a small ``blurring'' of the intrinsic proton charge, due to the motion
of the proton relative to the proton--$\pi^0$ center of mass, while the
neutron--$\pi^+$ configuration contributes to the charge distribution
at larger distances. This effect is most clear in the neutron electric form
factor, which would be extremely small in the absence of such effects, and
where the contribution from fluctuations into bound proton--$\pi^-$ yields a
positive core and a negative pion cloud.  A recent analysis~\cite{Fried}
showed suggestions of a similar structure in the proton form factors.

With precise new data in the low $Q^2$ region, 
is possible to better examine the form factors for indications of structure
related to the pion cloud and to improve extractions of the proton charge and
magnetic radii.  A series of low $Q^2$ polarization measurements~\cite{blast,
ron07, paolone10, ron11, zhan11}, the most recent of which achieved
$\approx$1\% total uncertainties on $\mugegm$ for $Q^2$ from 0.3 -- 0.7
GeV$^2$.  This is to be compared to typical uncertainties of 3--5\% from 
earlier cross section measurements, and $\gtorder$2\% for the previous
low-$Q^2$ polarization measurements. Figure~\ref{fig:gegm} shows the results
of these recent polarization experiments, which have dramatically improved the
precision of our knowledge of the form factor ratio $\mugegm$ at low $Q^2$. 
In addition, a new set of high-precision Rosenbluth measurements, focusing on
$Q^2$ values below 1~GeV$^2$, have been taken at Mainz~\cite{bernauer10} and
are in good agreement with the recent polarization data.  While the Rosenbluth
results are more sensitive to TPE contributions, TPE effects are generally
expected to be smaller, at the $\sim$1\% level, at these $Q^2$
values~\cite{borisyuk07, blunden05a}.  Nonetheless, these corrections can
still impact the extracted radii~\cite{blunden05b, arrington11c, bernauer11}.

Note that the new data on $\gegm$ rule out the kind of structure suggested
in the analysis of previous data~\cite{Fried}, although this does not
mean that the contribution from the pion cloud are small.  There could be
significant structure that cancels in the form factor ratio, or there could be
large pion cloud contributions which do not introduce any sort of ``bump'' or
other clear structure in the form factors.  The ratio $\mugegm$ is
significantly below unity over most of the measured range, with some
uncertainty at the lowest $Q^2$ values due to a small inconsistency between
the measurements in this region.  However, the more precise polarization data,
as well as the Mainz Rosenbluth results, indicate $\mugegm < 1$,
suggesting that the electric form factor falls off more rapidly than the
magnetic form factor at low $Q^2$.  If this holds down to $Q^2$=0, it implies
that the charge distribution is larger than the magnetization distribution. If
this is the case, then there is \textit{no energy scale at which the simple
non-relativistic picture, where the quark, charge, and magnetization
distributions are identical, is valid}. Even when probed at the lowest
momentum scales, we are sensitive to the quark dynamics and the impact of the
electric and magnetic fields generated by the motion of the
highly-relativistic quarks.

The new data also provide updated extractions of the proton charge and magnetic
RMS radii.  In the non-relativistic limit, $\ge = 1 - Q^2 \langle r^2 \rangle
/6 + ...$, allowing for the extraction of the mean-square radius, $\langle r^2
\rangle$, of the proton from the slope of the form factor at $Q^2=0$.
While the relativistic case yields corrections to this approximation, the
slope of the form factor is taken as the standard definition used in all
extractions of the proton charge radius.  The analysis combining polarization
data with previous cross section measurements yields a charge radius of
0.875(10)~fm~\cite{zhan11}, consistent with the Mainz extraction of
0.879(8)~fm~\cite{bernauer10} and the 2010 CODATA value of
0.878(5)~fm~\cite{mohr12}, which is mainly derived from the Lamb shift in
hydrogen. The finite size of the proton yields one of the largest corrections
to the atomic structure of hydrogen, allowing precise measurements of the Lamb
shift to be used as a measure of the proton charge radius.

\begin{figure}[htbp]
\includegraphics[width=.85\textwidth]{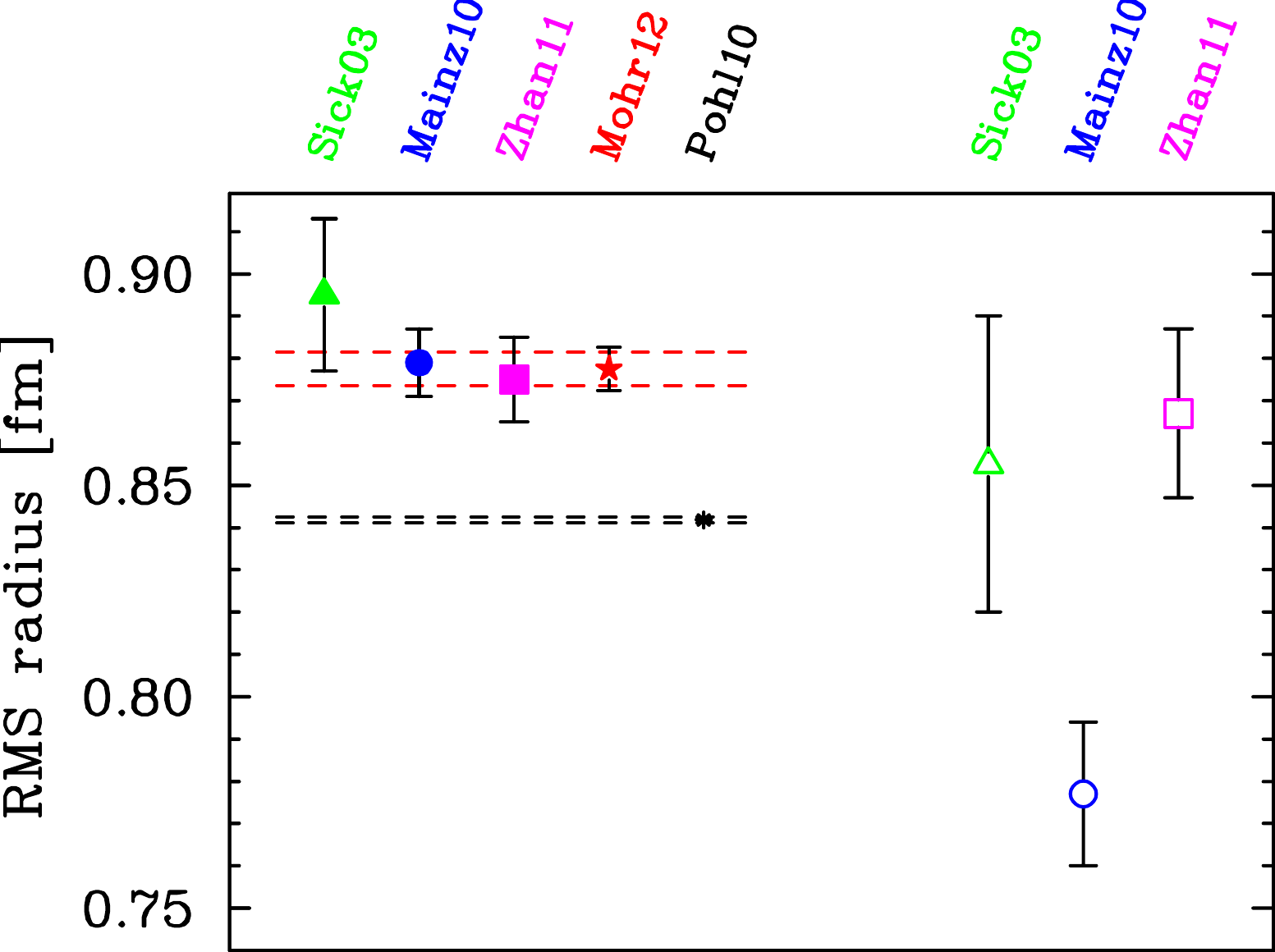}
\caption{The proton RMS charge radius (solid points) and magnetic radius
(hollow points) from electron scattering~\cite{sick03, bernauer10, zhan11} and
atomic physics~\cite{mohr12, pohl10} measurements. The red dashed lines show
the combined charge radius result from the electron-based CODATA, Bernauer,
and Zhan extractions (Sick is excluded as Zhan includes the same data), while
the black dashed lines show the Pohl uncertainty.}
\label{fig:radii}
\end{figure}

A recent measurement of the Lamb shift in muonic hydrogen~\cite{pohl10} yields
a proton RMS radius of $0.8418(07)$~fm, at odds with the electron scattering
results and the CODATA10~\cite{mohr12} value of $0.8775(51)$~fm. These
measurements of the proton charge radius are shown in Fig.~\ref{fig:radii},
The combined extraction based on electron--proton interactions are consistent,
and the combined result from these three measurements is $R_E$=0.8772(46),
9$\sigma$ above the muonic hydrogen result. The corrections applied to e--p
scattering and to the atomic hydrogen Lamb shift are very different, and the
agreement of the electron-based measurements suggests that the difference lies
in the muonic hydrogen result, either related to the measurement of the Lamb
shift, or the calculation of the QED corrections. Note that while there are
concerns about the effect of two-photon exchange in extraction of the
radius~\cite{blunden05b, ron11, arrington11c}, the impact of these corrections
appears to be small for the charge radius, but may account for some of the
discrepancy between the magnetic radii extracted in the electron scattering
measurements~\cite{bernauer11}.

With the use of new polarization techniques, our picture of the proton is
rapidly coming into much sharper focus, answering many questions about the
structure of the proton.  Recent high-$Q^2$ polarization measurements
demonstrated the possible importance of two-photon exchange corrections and found clear
indications of the key role of orbital angular momentum and quark dynamics in
the proton structure.  At low $Q^2$, our new polarization data show a clear
breakdown of non-relativistic quark models even at low momentum scales and
provide evidence of a difference in the size of the proton's charge and
magnetization distributions.

These measurements have raised new questions as well: are the two-photon
exchange corrections observed at high $Q^2$ also yielding differences between
new cross section and polarization measurements at very low momentum transfer,
and do they, or other corrections, explain the difference between various
extractions of the proton's charge radius. In the absence of an error in the
measurement or QED corrections, this significant discrepancy indicates that
something is missing in our description of the electromagnetic interaction
which yields a difference between electron and muon interactions. If the
difference between the proton radius obtained in e--p and $\mu$--p
interactions survives a more detailed examination, does this imply that
something is missing in our description of the hydrogen atom or in our
understanding of electron and muon interactions with matter?

This work was supported by the U.S. Department of Energy, Office of Nuclear
Physics, under contract DE-AC-06CH11357

\bibliographystyle{aipproc}   

\bibliography{440_Arrington}

\end{document}